\documentclass[aps,prl,twocolumn]{revtex4}
\usepackage{times}
\usepackage{latexsym}
\usepackage{graphicx}
\usepackage{verbatim,times,bbm}
\usepackage{color}
\usepackage{appendix}

\def\uno{\mbox{1 \kern-.59em {\rm l}}}

\def\be{\begin{equation}}
\def\ee{\end{equation}}
\def\ba{\begin{eqnarray}}
\def\ea{\end{eqnarray}}
\def\lo{\longrightarrow}

\def\hh{\hskip 2cm}
\def\la{\langle}
\def\ra{\rangle}

\begin{document}

\title{Quantum Secret Sharing and Random Hopping;\\
Using single states instead of entanglement}

\author{V. Karimipour, and M. Asoudeh}
\affiliation{ Department of Physics, Sharif University of Technology, Teheran, Iran}
\affiliation{ Department of Physics, Azad University, Northern Branch, Tehran, Iran}

\begin{abstract}
Quantum Secret Sharing (QSS) protocols between $N$ players, for sharing classical secrets, either use multi-partite entangled states or use sequential manipulation of single $d-$ level states only when $d$ is prime \cite{Tavak}. We propose a sequential scheme which is valid for any value of $d$.  In contrast to  \cite{Tavak} whose efficiency (number of valid rounds)  is $\frac{1}{d}$, the efficient of our scheme is $\frac{1}{2}$ for any $d$. This, together with the fact that in the limit $d\lo \infty$ the scheme can be implemented by continuous variable optical states, brings the scheme into the domain of present day technology.  
\end{abstract}

\pacs{03.67.Hk, 03.67.-a, 03.67.Dd}

\maketitle

{\bf Introduction:}
Quantum Key Distribution (QKD) \cite{BB84, Ekert} is among the best experimentally developed areas \cite{Gisin} in the field of quantum information science. With the rapid growth of demand for secure communication in our world,  it is imaginable that in the near future it will soon be an integrated part of  modern communication systems.  At the very basis of this developments lies the fact that QKD either uses bi-partite entanglement \cite{ Ekert} or sequential preparation and measurements of states by two parties, both of which are within reach of present technologies. 
In view of the need for more and more collaborative activities, Quantum Secret Sharing (QSS) is at least at the same level of demand as QKD. QSS is  the act of splitting a message so that only by collaboration of all the receivers, it can be retrieved. The question now is: is it possible to put Quantum Secret Sharing (QSS) between an arbitrary number of parties, on the same level of experimental feasibility as QKD? 
Until a few years ago, the answer seemed to be no, because the known QSS schemes relied on multi-partite entangled states which are much more difficult to prepare and more fragile than bi-partite ones.  Recent developments seem to point to a positive answer to this questions. \\

The idea of using entanglement for sharing a secret key between three persons, was first briefly pointed out in \cite{Zuk} and then developed in \cite{BHB}, where it was shown that local measurements of a  GHZ  state \cite{GHZ}, enable three or four parties, to share a random sequence of bits as a secret key. Generalization to many parties using multipartite entangled states were developed in \cite{SenZuk, Pan, ZhangLiMan}.  Generalization of QKD and QSS schemes to $d-$ level states, again relying on multipartite entanglement, were studied in \cite{CerfQKD} and \cite{YuLin} respectively.  Finally generalization of these entanglement-based schemes to continuous variable states \cite{CVReview}, were reported in many works \cite{Chen, Gaert, TycSanders, Lance1, Lance2, GrosshansNature, GrosshansPRL, CVQKD}, where in some of them like \cite{Chen, Gaert} the polarization degrees of freedom of photons and in the others coherent or Gaussian states of light were used. \\

Along with all this theoretical and experimental success and in view of the difficulty in creating and maintaining multi-party entangled states, a basic question has always been whether it is possible to do QSS between $n$ parties by using less amount of entanglement?  This  question  received a partial positive answer, where it was shown that multipartite entanglement of qubits can be replaced by two-body entanglement \cite{Karlsson} or by product states \cite{Guo, YanGao},   the later works being a combination of two or more QKD protocols of BB84 types \cite{BB84} and all of them using qubit states. A sequential scheme with a proof of principle for its experimental realization was then reported in \cite{Schmidt} where its security problem was discussed in  \cite{Qin} and \cite{He}.  \\

It was then quite natural to ask if this sequential scheme can be generalized for d-level states to which a positive answer was found for $d$ being an odd prime in \cite{Tavak}.  The efficiency ( the average percentage of valid rounds) of this scheme which uses a system of $d+1$ Gaussian MUB's is $\frac{1}{d}$, which becomes negligible for large $d$ and vanishes for continuous variables. It is the aim of this letter is to propose an entanglement-free QSS protocol which is valid for any $d$, uses the simple generalized Pauli and Hadamard gates and leads to familiar operations in the continuous variable limit and moreover whose efficiency is $\frac{1}{2}$ for all $d$. We hope that removing the previous restrictions, will bring the QSS scheme closer to experimental realization with current technology. \\

In order to proceed a quick review of the method of \cite{Tavak} is in order, which is based on using Mutually Unbiased Bases (MUB) in Hilbert spaces of prime dimensions. There are $N+1$ players which are denoted by $R_0$,  $R_1, R_2, \cdots R_{N}$.  
 The first player $R_0$ prepares a reference state and acts on it by an operator which depends on two random integers $0\leq a_0,b_0\leq d-1$ and passes the state to the next player who acts similarly on the state. In the end the state which reaches the last player $Bob$ depends on two parametes $A:=\sum_{i=0}^N a_i$ and $B:=\sum_{i=0}^{N} b_i$. This player always measures the received state in a fixed basis and only when $B=0$ which happens 1 out of $d$ runs, there is perfect correlation between his measured result $m$ and the random numbers $a_i$ in the form 
 
 \be\label{zukk}
 \sum_{i=0}^{N} a_i =m.
 \ee
 
In other rounds no correlation exist and the round is discarded.  If the protocol is run for $\approx Ld$ rounds and a number of $L$ rounds are successful, then a sequence of perfectly correlated random strings are shared between all the players. If we denote  by $K_i$ the long sequence of dits ($a_{i,1}, a_{i,2}, a_{i,3} \cdots a_{i,L}$) for the $i-$th player, and by ${\cal K}=(-m_1, -m_2, \cdots -m_L)$ the long sequence of negative-measured values by $R_N$, then according to (\ref{zukk}) we have
\be
{\cal K}\oplus K_0\oplus K_1 \oplus K_2 \oplus  \cdots K_{N}=0
\ee
where by $\oplus$ we mean  a dit-wise sum modulo $d$ between all the strings. \\

While this is an important step, the restriction to prime dimensions and the specific form of the operators used by the players make it difficult to go to the limit of continuous variables, a limit which has been most promising in actual implementation of many protocols by optical means \cite{CVReview}.  \\

In this letter we report an entanglement-free QSS scheme between $N+1$ players who use a   $d-$ level state, where $d$ is arbitrary. With $d$ being arbitrary, one can now hope that experimental implementation will be much more feasible.  

{\bf QSS scheme using a single qudit state for general $d$: }
The key insight is to look at the protocol of \cite{Tavak} as a random walk in a lattice of states, comprised of the bases of  MUB's. 
With this random walk picture at hand, we can alleviate the restriction to prime numbers by devising a new lattice and new hopping operators. We take a $4\times d$ lattice folded into a torus, where $d$ is arbitrary. 
Let $d$ be any positive integer and consider the computational orthonormal basis states for  $C^d$, the complex $d$ dimensional Hilbert space as: 
\be
{\cal B}=\{|k\ra, \ \ \  0\leq k \leq d-1\}
\ee
Let $\omega=e^{\frac{2\pi i}{d}}$ be the $d-$th root of unity and  $F$ be the Fourier transform operator defined as 
\be
F:=\frac{1}{\sqrt{d}}\sum_{i,j=0}^{d-1} \omega^{ij}|i\ra\la j|.
\ee
The operator $F$ has the following interesting properties, which result from the identity $\sum_{i=0}^{d-1} \omega^{r}=d\delta_{r,0}$:
\be
F^2 = \sum_{k=0}^{d-1} |-k\ra\la k|\ ,\hh F^4= I.
\ee
Furthermore, when acting on the state $|k\ra$, $F$ creates another state $|a_k\ra$
\be
|\xi_k\ra=\frac{1}{\sqrt{d}}\sum_{j=0}^{d-1}\omega^{jk}|k\ra.
\ee
which constitute another basis $\hat{{\cal B}}$, 
\be
\hat{{\cal B}}=\{|\xi_k\ra,\ 0\leq k\leq d-1\},
\ee
mutually unbiased basis  with respect to ${\cal B}$.
\begin{figure}[t]
 \centering
   \includegraphics[width=7cm,height=4.8cm,angle=0]{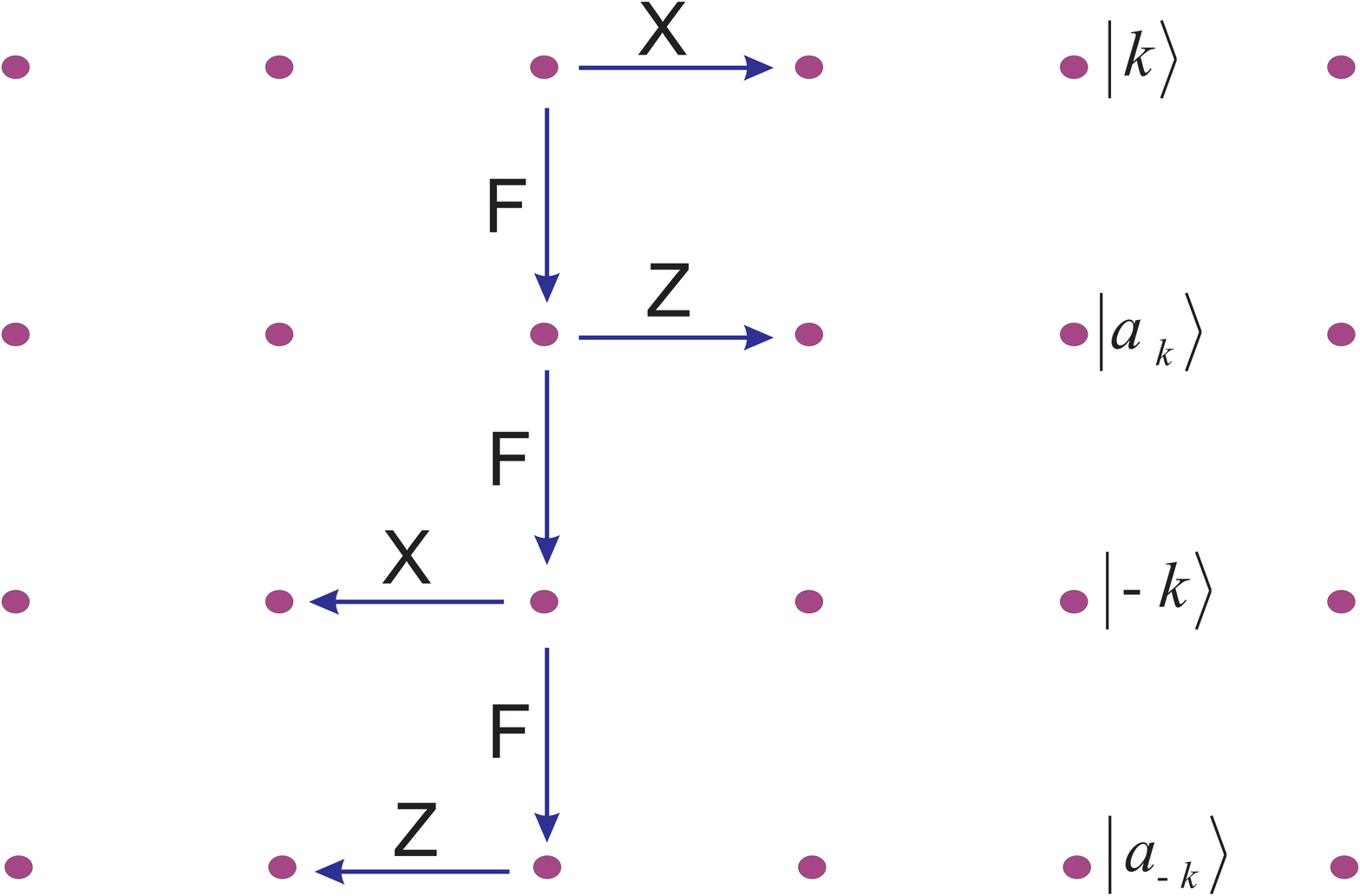}
   \caption{Color Online.
The action of each player causes the point representing the state to do a random hopping in  this lattice of states. It is important to note that the $Z$ operators on the first and the third rows and the $X$ operators on the second and fourth row do nothing (they just add overall phases). These phases can safely be ignored. 
    }
   \label{lattice2}
\end{figure}
Consider now the generalized Pauli operators $X:=\sum_{k=0}^{d-1}|k+1\ra\la k|$ and $Z:=\sum_{k=0}^{d-1}\omega^{k}|k\ra\la k|$. These two operators act as shift operators on the bases ${\cal B}$ and $\hat{{\cal B}}$ respectively:
\be
X|k\ra=|k+1\ra, \hh Z|\xi_{k}\ra=|\xi_{k+1}\ra. 
\ee
Conversely when acting on the bases $\hat{{\cal B}}$ and ${\cal B}$, the operators $X$ and $Z$ act as phase operators:
\be
X|\xi_k\ra=\omega^{-k}|\xi_k\ra,\hh Z|k\ra = \omega^k |k\ra. 
\ee
We also note that:
\ba
F|k\ra=|\xi_k\ra, \hh F|\xi_k\ra=|-k\ra.
\ea
In view of these relations it is convenient to draw the four rows of states of ${\cal B}$ and $\hat{{\cal B}}$ in the order shown in figure (\ref{lattice2}). \\

{\bf Running the protocol:}

\begin{figure}[t]
 \centering
   \includegraphics[width=7cm,height=4.8cm,angle=0]{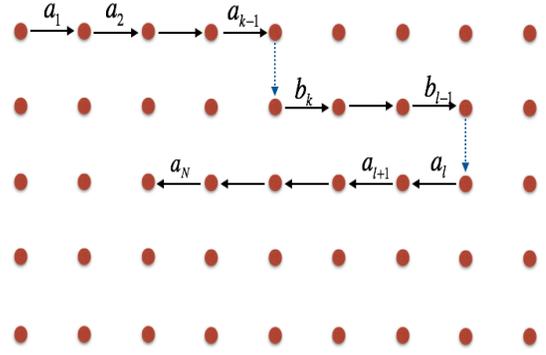}
   \caption{Color Online.
A simple example of a random path in the lattice of states, when only 2 players apply $F$ operators.  In order not to clutter the figure, we have shown all the steps to be of unit-size (between neighboring sites). The steps can be of any size between non-neighboring sites. The first and the last rows are the same.
    }
   \label{Path1}
\end{figure}

\begin{figure}[t]
 \centering
   \includegraphics[width=7cm,height=4.8cm,angle=0]{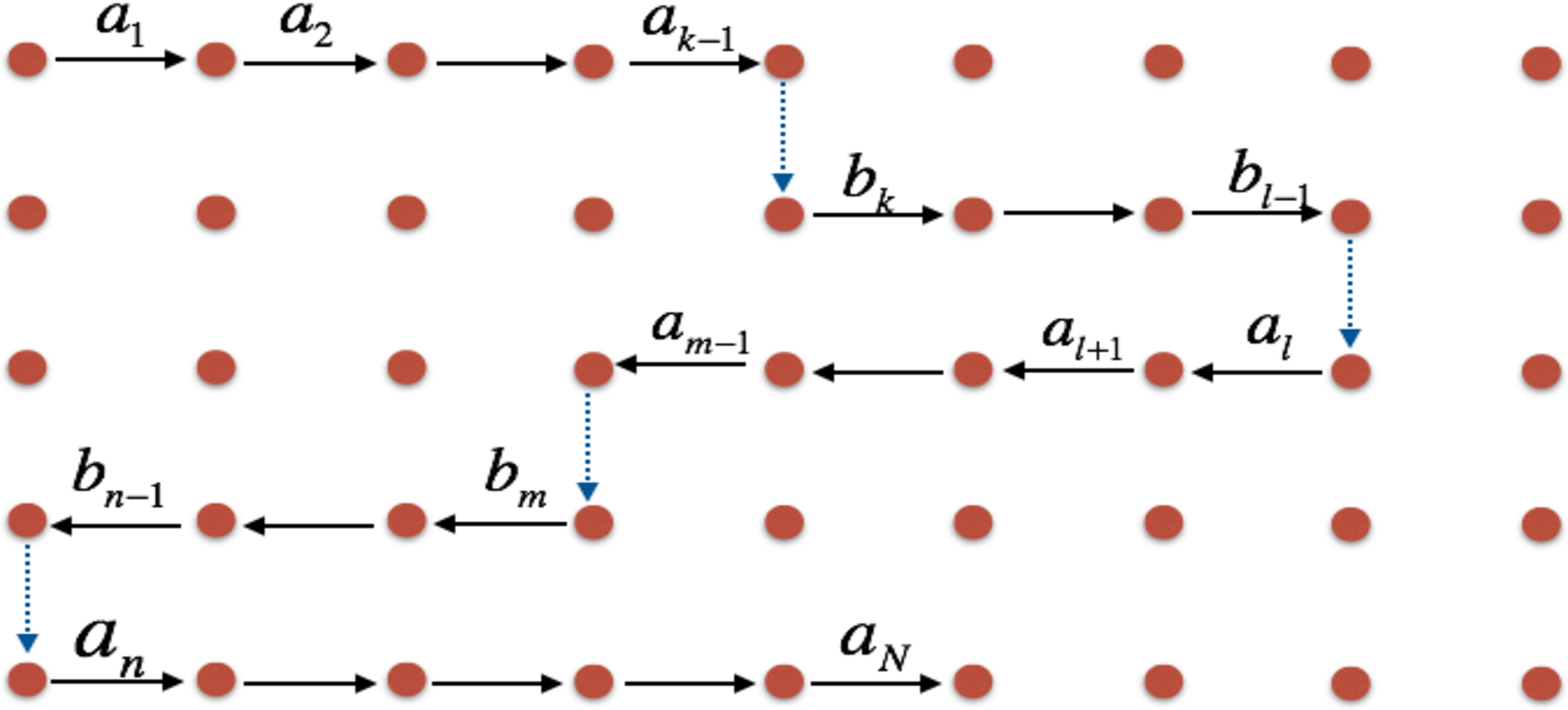}
   \caption{Color Online.
A simple example of a random path in the lattice of states, when only 4  players apply $F$ operators.    }
   \label{path2}
\end{figure}

\begin{itemize}
\item{\bf Step 1:}   $R_0$, starts from the state $|0\ra$ and acts on it by the operators $X^{a_0}Z^{b_0}F^{c_0}$ (where $0\leq a_0, b_0 \leq d-1$ are random integers and $c_0=0, 1$). She then sends this state to the first player $R_1$ who acts on the received state by $X^{a_1}Z^{b_1}F^{c_1}$ and then sends the state to the second player $R_2$ who does the same thing. The process repeats until the state reaches the last player  $R_N$. He also acts on the state by  
$X^{a_N}Z^{b_N}F^{c_N}$ and measures it in the computational or ${\cal B}$ basis. The final state before this measurement is given by 
\be
|\Psi\ra=\prod_{i=0}^N (X^{a_i}Z^{b_i}F^{c_i})|0\ra.
\ee
\item{\bf Step 2:}
 After measurement of $R_N$, all the players $R_0, R_1, R_2 , \cdots R_{N}$ are asked by Alice (the one who controls the protocol whom we call Alice. She can be any of the players) to publicly announce their integers $c_i$ which are $0$ or $1$.  Also they can be asked by Alice to make this announcement in random order.  It is crucial that the integers $a_i$ and $b_i$ are kept secret with the players and are not announced at any stage. If    
\be
\sum_{i=0}^N c_i=0 \hh {\rm mod} \ 2
\ee
the round is treated as valid, since this means that the point has landed on the correct basis ${\cal B}$ for the measurement of $R_N$,  otherwise it is discarded.  \\
\item{\bf Step 3:}
In  valid rounds, we are certain that the point representing the state lies on the computational basis $B$ and when $R_N$ measures this state he obtains a definite and deterministic value denoted by $m_N$.  In invalid rounds, due to the MUB property of the two bases $B$ and $\tilde{B}$, a completely random result is obtained and no perfect correlation exists between the result of $R_N$ measurements and the random bits applied by others. \\
  \end{itemize} 
But what is the exact form of the correlation?  It is not as simple as (\ref{zukk}), but it is as perfect as it is. Looking at the lattice in figure (\ref{lattice2}), it is crucial to note that on the first and third rows, the $Z$ operator does nothing (it just adds overall phase to the state). The same is true for the $X$ operator on the second and fourth rows. This means that the effect of these operators on these  respective rows can safely be ignored.  So suppose that the operator  $F$ is applied only twice, by the players $R_{k}$ and $R_l$. Figure (\ref{Path1}) shows the path of the random walk. Modulo an overall phase, the final state is given by:
 \be\label{akk1}
 |\Psi\ra=|\sum_{i=0}^{k-1}a_i + \sum_{i=k}^{l-1} b_i - \sum_{i=l}^N a_i\ra:=|A_{0,k-1}+B_{k,l-1}-A_{l,N}\ra,
 \ee 
where
\be
A_{r,s}:=\sum_{i=r}^{s-1} a_i,\hh B_{r,s}:=\sum_{i=r}^{s-1} b_i. 
\ee
Or suppose that the operator $F$ is applied four times, by the players $R_k$, $R_l$, $R_m$ and $R_n$. Figure (\ref{path2}) shows the path of the random walk and the final state is given by:
\be
|\Psi\ra=|A_{0,k-1}+B_{k,l-1}-A_{l,m-1}-B_{m,n-1}+A_{n,N}\ra
\ee
 The pattern is now clear. Although it is straightforward, we do not attempt to clutter the text with writing general formulas for the final state corresponding to a given random path. The point to emphasize is that once the numbers $c_i$ are publicly announced, all the parties; $R_0$ to $R_N$ know the positions of the $F$ operators, the vertical steps. Note that the vertical steps are of unit size.  In those valid   
 rounds, the value measured by $R_N$  has perfect correlation with those used by  all the other parties. Let us elaborate this by an example. Suppose there are 6 players $R_0$ to $R_5$ and two $F$ operators are used in round 1 and round 2, but in two different places. For example in round 1 the F operators are in positions (2 and 5) and in round 2 the F operators are in positions (1 and 4).    Then according to (\ref{akk1}) if we denote the values that $R_5$ measures in these two rounds respectively by  $m$ and $m'$, we will have 
 \ba
 a_0+a_1+b_2+b_3+b_4-a_5&=&m_5\cr
 a'_0+b'_1+b'_2+b'_3-a'_4-a'_5&=&m'_5.
 \ea
 
Since the positins of the $F$ operators (the vertical steps) are known by the public announcement of all the $c_i$'s,  they know how to arrange their Keys $K_0, K_1, \cdots K_5$ from the above $d-$ level integers. For example,  the first numbers of these Keys are:
 $
 K_0 = (a_0, a'_0,\cdots), \ K_1 = (a_1, b'_1,\cdots)\  K_2 = (b_2, b'_2,\cdots), \
  K_3 = (b_3, b'_3,\cdots), \ 
 K_4 = (b_4, -a'_4,\cdots),\
 K_5 = (-a_5, -a'_5,\cdots), $ and $
 {\cal M}_5= (-m_5, -m'_5,\cdots).
 $
 Then we have 
 \be
 K_0 \oplus K_1 \oplus K_2 \oplus K_3 \oplus K_4 \oplus K_5 \oplus {\cal M}_5=0 .
 \ee
  In this way a secret key is created between all the parties which can be used for sharing messages via public classical channels. 
  \begin{itemize}
  \item{{\bf Using the shared key:}} Let us denote the random string $K_N+{\cal M}_N$ of the last player by ${\cal K}_N$. Therefore the shared random key which has been established satisfies: $K_0 + K_1 + \cdots K_{N-1}+{\cal K}_N=0$
    
Like any other QSS scheme, any player $R_i$ can play the role of Alice who wants to send a secret message to another player $R_j$ (playing the role of Bob) who will retrieve the message by the collaboration of other parties.  Alice can send a message ${\cal M}$ in the form ${\cal M}\oplus K_i$ to Bob.   Alice who now controls the protocol, (like any other QSS scheme) asks the other parties to send their random keys to Bob who retrieves the message by performing the relation $({\cal  M}\oplus K_i)\oplus \sum_{j\ne i} K_j $ which in view of the previous relation gives the message ${\cal M}$. ‌\\
 \end{itemize}
 
{\bf Security against attacks:}
We can now consider how the protocol is secure against attacks. A round of the protocol corresponds to a random path on the lattice of states as shown in figure (\ref{path2}).  If any of the players wants to measure his  received qudit and keeps a copy of it, he should know the basis, but this is not possible for him, because the integers $c_i$ or equivalently the $F$ moves, are announced only after all measurements are done. Since the two bases ${\cal B}$ and $\hat{{\cal B}}$ are MUB with respect to each other, blind measurement in a basis, causes an error rate of $\frac{d-1}{d}$ in the final measured qudit by Bob. This type of intervention or cheating is easily detectable by publicly comparing a subsequence of the shared key.  These considerations also apply if a group of players enter a plot to intercept the key, since they can be collectively considered as a single party to which the basis used by the previous parties is not known.  \\

Another conceivable attach is that a group of players, each entangle their respective received qudits from  previous players to some ancilla and  store part of the information in these ancillas. They can then collaborate with each other and share this collected information to retrieve the key. A standard way for entangling a received state $|\psi\ra$ to an ancilla is to start the ancilla in the state $|+\ra:=\frac{1}{\sqrt{d}}\sum_{i=0}^{d-1}|i\ra$   and act on the joint state $|+\ra\otimes |\psi\ra$ by a (generalized Controlled Not (CNOT) operator,  
$CNOT|i,j\ra:=|i,i+j\ra$. If the received state is a computational state $|q\ra$, , then this action results in 
$
CNOT|+\ra\otimes |q\ra=\frac{1}{\sqrt{d}}\sum_{i=0}^{d-1} |i\ra\otimes |i+q\ra,
$
which is a maximally entangled state. In this way the ancilla will be in a completely mixed state and stores no information of the qudit $q$. However if the received qudit is of the form $|\xi_q\ra$, that is, a Fourier basis state, then the ancilla can indeed store 
some information of $q$. In this case we find  that
$
CNOT|+\ra\otimes |\xi_q\ra=|\xi_{-q}\ra\otimes |\xi_{q}\ra.
$
In this  way, information about $q$ is stored in the ancilla without leaving any trace in the state $|\xi_q\ra$ which is being communicated between the players. Each player can then measure his ancilla in the Fourier basis $\hat{B}$ and retrieve $q$. However since a received state can be in one of the states $|q\ra$ or $|\xi_q\ra$ with equal probability, each player has a $50$ percent chance of successfully storing  information in his ancilla, leaving no trace in the transmitted state $|\xi_q\ra$, and $50$ percent chance of completely de-cohering the state $|q\ra$. In this second case he commits an error rate of $\frac{d-1}{d}$ on each state. Again by publicly announcing a subsequence of the key, the players can discern the existence of a plot within the group.\\

{\bf Conclusion}
We have proposed a new scheme for multi-party secret sharing which uses a single $d-$ level state for arbitrary $d$,  and alleviates the need for entanglement.  The basic idea is the analogy with a random walk performed by a particle through a lattice of states. This walk is the result of random sequential unitary operations on a single particle by the players one after the other. In view of the fact that the players do not have information about  each others unitary operations, the path of the particle resembles a random walk which lands on a particular row (a particular basis) only when certain conditions are met. This condition is revealed only after public announcement of some parameters of the operations after all the measurements have been done. 
In the limit of $d\lo \infty$, the local actions of the players will become $X(s)Z(t)F^{c}$ where $s$ and $t$ are continuous variables and $c=0,1$. Here $X(s)=e^{is\hat{p}}$ and $Z(t)=e^{it\hat{x}}$ with $\hat{x}$ and $\hat{p}$ the position and momentum operators ($[\hat{x},\hat{p}]=i\hbar$) (corresponding to the quadratures of aa mode of electromagnetic field) and $F$ the Fourier transform between the position and momentum bases. Quantum optical realizations of these relations then puts this protocol quite within the reach of present day technology.

\end{document}